\def\ts     {\thinspace}
\def\,      {\thinspace}
\def\kms    {\ifmmode{{\rm \ts km\ts s}^{-1}}\else{\ts km\ts s$^{-1}$}\fi}

\def\msun   {$M\rm_\odot$}

\def\lsun   {$L\rm_\odot$}
\def\lmsun  {$L_{\rm \odot}/M_{\rm \odot}$}
\def\lmvir  {$L_{\rm IR}/M_{\rm VIR}$}
\def\lm  {$L_{\rm IR}/M$}

\def\cs    {\ifmmode{{\rm CS}\ts J\!=\!2\! \to \!1}\else{{\rm CS}\ts $J$=2$\to$1}\fi}
\def\ppm   {$\pm$}
\def\hii   {H\footnotesize\,II\normalsize}

\documentclass{aastex61}

\usepackage{amsmath}
\usepackage{rotating}

\usepackage[switch]{lineno}

\shorttitle{Search for High-Mass Protostellar Objects in Cold IRAS Sources}
\shortauthors{Ao et al.}

\begin{document}


\title{Search for High-Mass Protostellar Objects in Cold IRAS Sources}

\author{Y. Ao}
\affiliation{Purple Mountain Observatory \& Key Laboratory for Radio Astronomy, Chinese Academy of Sciences, 8 Yuanhua Road, Nanjing 210034, China}
\affiliation{National Astronomical Observatory of Japan, 2-21-1 Osawa, Mitaka, Tokyo 181-8588, Japan}

\author{J. Yang}
\affiliation{Purple Mountain Observatory \& Key Laboratory for Radio Astronomy, Chinese Academy of Sciences, 8 Yuanhua Road, Nanjing 210034, China}

\author{K. Tatematsu}
\affiliation{National Astronomical Observatory of Japan, 2-21-1 Osawa, Mitaka, Tokyo 181-8588, Japan}

\author{C. Henkel}
\affiliation{MPIfR, Auf dem H\"{u}gel 69, 53121 Bonn, Germany}
\affiliation{Astron. Dept., King Abdulaziz Univ., P.O. Box 80203, Jeddah 21589, Saudi Arabia}

\author{K. Sunada}
\affiliation{Mizusawa VLBI Observatory, NAOJ, 2-12 Hoshi-ga-oka, Mizusawa-ku, Oshu-shi, Iwate 023-0861, Japan}

\author{Q. Nguyen-Luong}
\affiliation{National Astronomical Observatory of Japan, 2-21-1 Osawa, Mitaka, Tokyo 181-8588, Japan}
\affiliation{Korea Astronomy and Space Science Institute, 776 Daedeokdae-ro, Yuseong-gu, Daejeon 34055, Korea}

\begin{abstract}
We present the results of \cs\, mapping observations towards 39 massive star
	forming regions selected from the previous CO line survey of cold IRAS
	sources with high-velocity CO flows along the Galactic plane (Yang et
	al. 2002).  All sources are detected in \cs, showing the existence of
	CS clumps around the IRAS sources. However, one third of the sources
	are not deeply embedded in the dense clumps by comparison of the
	central powering IRAS sources and the morphologies of CS clumps.
	Physical parameters of the dense molecular clumps are presented. We
	have identified 12 high-mass protostellar object (HMPO) candidates by
	checking the association between the dense cores and the IRAS sources,
	the detection of water maser, and the radio properties towards the IRAS
	sources. We find that the HMPO sources are characterized by low FIR
	luminosity to virial mass ratios since they are in very early
	evolutionary stages when the massive protostars have not reached their
	full luminosities, which are typical for zero-age main sequence
	stars.  Large turbulent motion in the HMPO sources may be
	largely due to the large kinetic energy ejected by the central
	protostars formed in the dense clumps. However, alternative means or undetected
	outflows may also be responsible for the turbulence in the clumps.

\end{abstract}


\section{Introduction}

In recent years, more attention has been paid to the study of massive star
formation (MSF), due to their predominant role in the evolution of the
interstellar medium and the Galaxy. To study the physical properties,
especially the initial conditions, of MSF, it is very important to search for
high-mass protostellar objects (HMPOs) before the energetic processes such as
powerful jets, outflows and strong ultraviolet (UV) radiation destroy their
natal clouds in the massive star forming regions (MSFRs). HMPOs are usually
formed in clusters and tend to be deeply embedded in the gas clouds.  At the
early stage of MSF, a HMPO can not produce a detectable H\footnotesize\,II
\normalsize region, even though the central protostar produces a lot of UV
photons, because these photons can not travel far from the protostar before
being absorbed by the surrounding material and the size of the ionized gas
region is extremely small, making it undetectable at centimeter wavelengths
with current sensitivities (Churchwell 2002). Therefore, HMPOs are
characterized by high infrared luminosity, strong dust emission and very weak
or no detectable free-free emission at centimeter wavelengths (Garay \&
Lizano 1999; Hosokawa \& Omukai 2009; Hosokawa et al. 2010).  As
a consequence, large surveys aiming at characterizing the
physical properties of MSFRs and searching for embedded massive young stellar
objects (YSOs) have been made using high-density tracers, e.g., CS, HCO$^{+}$,
or HCN, and submillimeter dust continuum emission (Plume et al.  1992, 1997;
Zinchenko et al.  1994, 1995, 1998; Bronfman et al. 1996; Juvela 1996; Molinari
et al.  1998a; Cesaroni et al.  1999; Hunter et al. 2000; Brand et al. 2001;
Sridharan et al. 2002; Beuther et al. 2002a; Shirley et al. 2003; Ao et al.
2004; Fa\'{u}ndez et al. 2004; Williams et al. 2004; Fontani et al.  2005;
Klein et al. 2005; Schnee \& Carpenter 2009; Elia et al. 2017).
In particularlly, the Red MSX Source (RMS) survey is a large MSX-selected
sample with multiwavelength data (Urquhart et al. 2007a,b, 2008, 2009a,b,
2011a,b, 2012, 2014a, 2015; Mottram et al. 2007, 2010, 2011; Cooper et al.
2013; Lumsden et al. 2013; Maud et al. 2015a,b). The APEX Telescope Large Area
Survey of the Galaxy (ATLASGAL) survey (Schuller et al. 2009) is the first
unbiased continuum survey of the whole inner Galactic disk at 870$\mu$m
(Tackenberg et al. 2012; Wienen et al. 2012, 2015; Urquhart et al.  2013a,b,
2014b; Csengeri et al.  2016a,b, 2017; Li et al. 2016; Giannetti et al.  2017;
Tang et al. 2018; Yang et al. 2018). Both RMS and ATLASGAL surveys are ideal
samples for identifying YSOs and are useful in comparison with other
studies.

The search for HMPOs in the Milky Way was often based on source
catalogs from the unbiased all-sky infrared surveys and color-color selection
criteria (Wood \& Churchwell 1989, Yang et al. 2002). Further removal of
contamination induced by color-color selection is conducted by detecting dense
gas tracers such as CS, HCO$^{+}$, and HCN and early-stage star formation
activity such as masers in the HMPO candidates (Beuther et al. 2002a; Wu et al.
2010).  While the new all-sky infrared compact source catalog from the Planck
mission is available now (Planck Consortium et al. 2016), the older all-sky
infrared source catalog from IRAS satellite is still a good foundation to
search for HMPOs in the Milky Way as the peaks of the HMPOs spectral energy
distribution is at the IRAS wavebands of 60--100\,$\mu$n (Sridharan et al.
2002). Yang et al. (2002) applied cold-color selection criteria to the IRAS
source catalog to search for CO in 1912 cold IRAS sources belonging to the
northern Galactic plane.  1331 of these sources were detected. In that paper,
they also identified 289 high-velocity CO flows (HVFs) on the basis of the
existence of broad CO line wing emission. Among these HVFs, 151 sources have an
infrared (IR) luminosity $>$\,10$^3$\,\lsun\ and are thought to be a good
target sample of early stage MSFRs.  Although the central massive stars have
been formed around these IRAS sources, the natal molecular clouds must still
exist and can provide essential clues for ongoing or incipient massive star
formation.  In this paper, we conduct a search for HMPOs by randomly selecting
39 out of the 151 luminous HVFs (Yang et al. 2002) and perform CS 2--1 mapping
with the Nobeyama 45m radio telescope. It is an extension of our
previous work on the \cs\, study of 10 IRAS sources (Ao et al.  2004). We
describe the observations in Section\,\ref{observation}. The observational
results and the HMPOs candidates are discussed in Section\,\ref{results}. In
Section\,\ref{discussion}, we discuss the physical properties of the HMPO
sample focusing on the luminosity to mass ratio and its implication to
massive star formation. Finally, we provide our conclusion in
Section\,\ref{conclusion}.

\section{Observations}\label{observation}
Observations were carried out with the 45m telescope of the Nobeyama Radio
Observatory (NRO) in 2004 February 21 to 23 and 2005 May 21 to 23.  The
receiver front end was the 25 BEams Array Receiver System (BEARS) (Sunada et
al. 2000). The half-power beam width of the NRO 45m telescope was approximately
17$\arcsec$ for the \cs\, observations.  The double sideband system temperature
was around 200-400 K, depending on weather conditions and different beams of the
BEARS.  The main beam efficiency was about 0.52 at 100 GHz and the velocity
resolution was about 0.1\,\kms.

The observations were performed in a snapshot mode. Since the beam separation
of the BEARS was 41.1$\arcsec$, four snapshots were observed to reduce the grid
spacing to one half of the beam separation, i.e. 20.5$\arcsec$, which is
slightly larger than the beam size of 17$\arcsec$. Figure~\ref{profile} shows
the mapping positions and profile map towards one source, IRAS~02232+6138.
Spectra marked by a cross symbol with the specific color were taken from one
snapshot observation, providing 25 spectra by the BEARS simultaneously, and in
total 100 spectra were obtained. 10$\times$10 mapping observations were carried
out for most sources except for 04324+5106, 05375+3540, 18592+0108, 22543+6145
and 23140+6121. The mapping area of the latter five sources were extended to
cover the whole emission regions. The typical on-source time was about 10 to 20
minutes for one source. To obtain the scaling factor of each beam of the BEARS,
the calibrator S140 was observed with the BEARS and a single-sideband SIS
receiver S100. The r.m.s. noise temperature on a main beam brightness
temperature scale was about 0.6 K at a velocity resolution of 0.1\,\kms. The
pointing was checked by nearby SiO masers around the target every two hours
and the pointing accuracy was within 5$\arcsec$.

The source list is given in Table\,\ref{list}.  We give the IR luminosity of
associated IRAS point source calculated as in Casoli et al. (1986). The results
derived from the IRAS data are also presented in Table\,\ref{list}.

\section{Results}\label{results}
\subsection{Determining the distances of IRAS sources}
Distances were obtained from extensive literature search. For 10 sources, we
adopt their parallax measurements (for details see
Table\,\ref{list}). We take photometric distances for 16 sources associated
with OB associations and H\footnotesize\,II \normalsize regions when available.
For the remaining 13 sources, we calculate kinematic distances using the
revised prescription given in Reid et al. (2009). Of these, IRAS 18567+0700,
IRAS 20216+4107 and IRAS 20220+3728 have distance ambiguity.

{\noindent \bf IRAS 18567+0700: } If we adopt its far kinematic distance of 10.8
kpc, its infrared luminosity will be $\sim$1.4$\times$10$^5$~\lsun\, and should
be detectable at radio. However, it was not detected at 1.4 GHz with a 5$\sigma$
upper limit of 2.5~mJy. Thus, we adopt its near kinematic distance of 2.1
kpc.

{\noindent \bf IRAS 20216+4107: } The source is associated with Cygnus X complex, thus we
adopt its far kinematic distance of 2.3 kpc.

{\noindent \bf IRAS 20220+3728: } If we adopt the near kinematic distance of 0.33 kpc, its
infrared luminosity will be $\sim$3.2$\times$10$^2$~\lsun\, and the source is
unlikely to produce a detectable \hii\, region. However, this source is detected
with a radio flux of 0.63 Jy at 1.4 GHz. Thus, we adopt its far kinematic
distance of 3.7 kpc.

The typical uncertainties of the adopted distances are about 10 to 20\%. For
example, the after-mentioned \lmvir\, ratio in $\S$~\ref{hmpo-ratio} is
directly proportional to the distance. The variance at a level of 10 to 20\% will
not change our analysis and conclusions in the following.

\subsection{\cs\, results}
\begin{figure*}[t]
\centering
  \includegraphics[angle=0,width=1.0\textwidth]{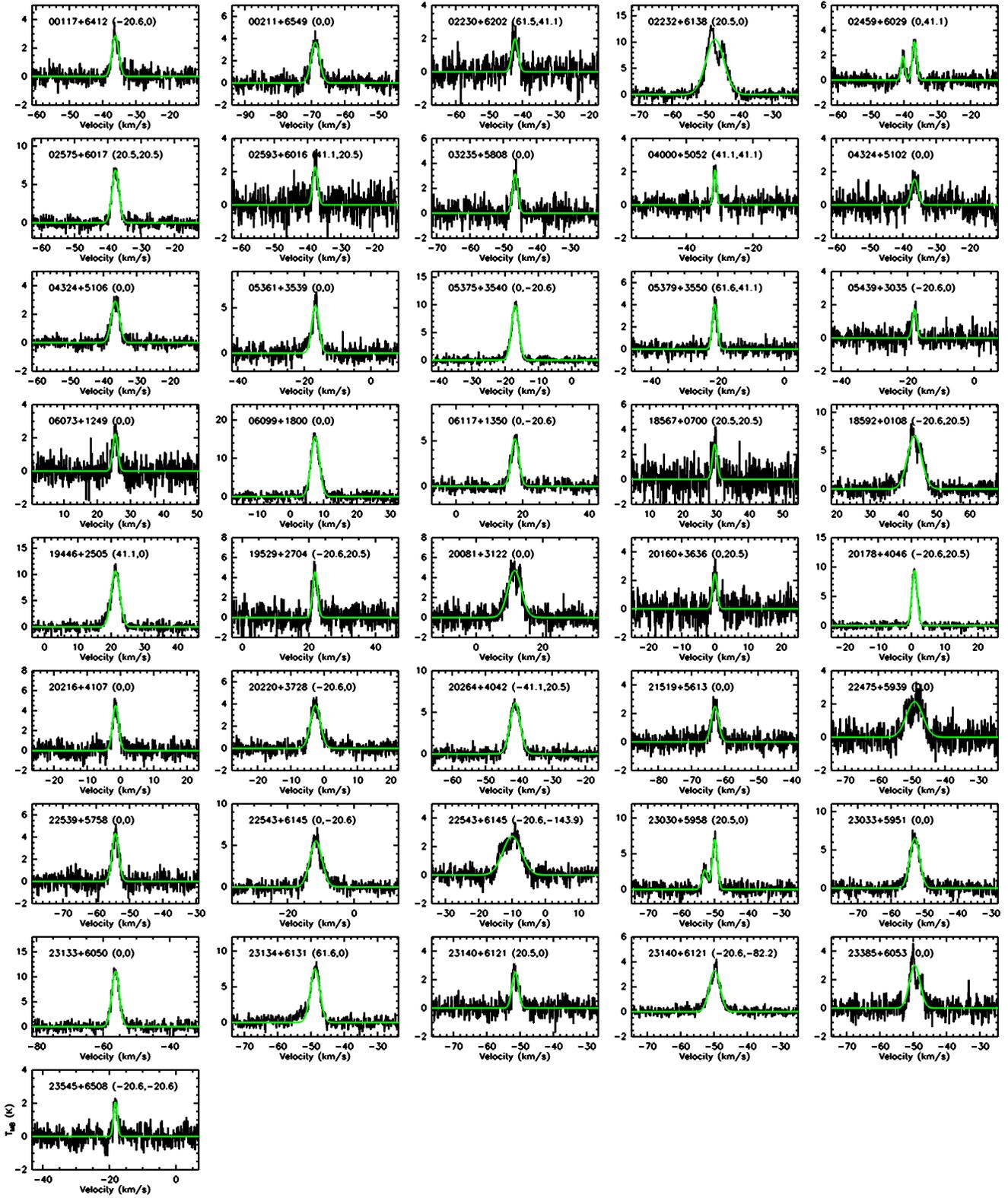}
  \caption{Spectra of the peak positions of clumps closed to the IRAS sources.
}\label{spec}
\end{figure*}

\cs\, emission has been detected towards all sources, showing the existence of
CS clumps around the IRAS sources. Figure~\ref{spec} shows the spectra of the
peak positions of CS clumps close to the IRAS sources. A few targets show
double peaked line profiles. IRAS~02459+6029 and IRAS~23030+5958 have two
distinct velocity components, which are associated with two distinct clumps.
Another two sources, IRAS~02232+6138 and IRAS~23385+6053, with blue-skewed
profiles may be caused by infalling motion of gas. The results, including
integrated intensities and positions of the peak spectra of the clumps, are
presented in Table\,\ref{spectable}. The peak intensities of the sources cover
a range from 1.64 K to 16.65 K, and the median and averaged values are 4.14 K
and 5.16 K, respectively.

Contour maps of \cs\, integrated intensity are shown in Figure~\ref{csmap}.
The typical mapping sizes are 200$\arcsec\times$200$\arcsec$, but for
IRAS\,04324+5106, 05375+3540, 18592+0108, 22543+6145 and 23140+6121, larger
maps are obtained to cover the whole emission regions. Physical parameters of
the clumps, such as radius $R$, averaged line width $\rm \overline{\Delta v}$
and virial mass $M_{\rm VIR}$, are calculated in the following ways. The angular extent
of each source is determined by finding the area within the half-power contour,
$A\rm_{1/2}$ and calculating the angular size of a circle with the same area.
The nominal radius, $R$, is determined by deconvolving the telescope beam,
$\theta_{\rm MB}$, via the formula $R$ = $D(\frac{A_{\rm
1/2}}{\pi}-\frac{\theta^{\rm 2}_{\rm MB}}{4})^{\rm 1/2}$, where $D$ is the
distance to the source.  The virial mass for a homogeneous spherical clump,
neglecting contributions from the magnetic field and surface pressure, can be
calculated from the expression $M_{\rm VIR}(M_{\rm \odot})$ = $0.509D({\rm
kpc})\theta({\rm arcsec})\rm \overline{\Delta v}^2$ where $\theta$ is the
deconvolved angular size and $\rm \overline{\Delta v}$ is the mean line width
of a positions within the \cs\, half-power contour. The results are presented
in Table\,\ref{core}. Clump radius, averaged line width and virial mass are
given in Column\,2-4. The typical radial size in the sample is 0.33\,pc,
ranging from 0.15\,pc to 1.66\,pc. The median values for the averaged line
width $\rm \overline{\Delta v}$ and virial mass $M_{\rm VIR}$ are 2.28\,\kms\,
and 360\,\msun, respectively. 

\subsection{HMPO candidates} 
To study the initial conditions of MSFRs at the early stage, we will identify a
sample of HMPO candidates by checking the association between the dense cores
and the IRAS sources, the detection of water masers, and the radio properties
towards the IRAS sources.

\subsubsection{Association between dense clumps and the IRAS sources}\label{dense}
Our sample is selected from IRAS sources with large far-infrared (FIR)
luminosities and broad CO line wing emission. Large luminosities suggest
massive young stars have been formed or are being formed at the sites of the
IRAS sources. The broad CO wing emission is likely caused by strong
winds/outflows from IRAS sources, suggesting the objects are still in an early
evolutionary stage. Thus our sample tends to define a collection of sources in
an early evolutionary stage. However, a few of these sources may already be
more evolved, where strong feedbacks have destroyed most of their natal
molecular clumps. To search for the HMPOs in the early stage, we should exclude
such sources by checking whether the IRAS sources are still deeply embedded in
their molecular dense clumps.

If an IRAS source is located within the \cs\, half-power contour, we classify
it as an embedded source within the dense molecular clump. In total, there are
26 sources embedded in the dense gas clouds while the remaining 13 sources are
offset from the dense gas peak (see Figure\,\ref{csmap} and Column\,6 of
Table\,\ref{core}).

\subsubsection{Radio continuum emission}\label{radio}
We used the radio data from the NRAO VLA Sky Survey (Condon et al. 1998), which
is a 1.4 GHz continuum survey covering the entire sky north of $-$40 deg
declination. We searched for the radio emission within 30$\arcsec$ of the
positions of the IRAS sources and their integrated flux densities are given
in Table\,\ref{hmpo-table}. For undetected sources, an upper 3$\sigma$ limit is
given. Then we calculate the geometry-independent parameters (the excitation
parameters and the Lyman continuum photon fluxes) as in Kurtz et al. (1994).
The results are summarized in Table\,\ref{hmpo-table}. Radio spectral types
of the stars (Panagia 1973) required to produce the Lyman continuum photon
fluxes, $N{\rm _{c}^{'}}$, of the sources are given in column 7 of 
Table\,\ref{hmpo-table}.

From the IR luminosity, one can estimate a stellar spectral type under the
assumption that the entire IR emission is due to a single star (Panagia 1973).
However, the single-star assumption could be too simple and not appropriate,
and we should assume that an IRAS source is the site where a star cluster is
formed. Therefore we adopt the universal initial mass function (IMF) of the
stars following Kroupa (2001) and the mass-luminosity relationship of the zero
age main sequence stars (Schaller et al. 1992). Then we can estimate the most
massive stars formed in the clusters with the comparable infrared luminosities
as the IRAS sources. The estimated IR spectral types of the most massive stars
are presented in Column~4 of Table\,\ref{hmpo-table}. We consider that a source
can not be a good HMPO candidate when its IR spectral type is later than the
radio spectral type derived as above and the difference between both spectral
types is more then one spectral type. This criterion will select the sources
without much evolved \hii\, regions compared to their FIR luminosities. This
criterion will exclude 22475+5939 and 23030+5958 as good HMPO candidates even
if they satisfy other criteria. The results are summarized in
Table\,\ref{hmpo-table}.

\subsubsection{Water masers}\label{water}
In star-forming regions, water maser emission at 22\,GHz is often associated
with high mass YSOs (Elitzur et al. 1989) and believed to be one
of the best tracers to investigate the earliest phases of star formation.
Interferometric observations suggest that water masers are associated with
jets or winds near the base of the outflows (Torrelles et al. 2003; Goddi et
al.  2005; Moscadelli et al.  2005).  All sources in our sample have been
observed to search for water masers (Sunada et al. 2007). The results are given
in Table\,\ref{hmpo-table}. Water masers are detected towards 25 out of 39
sources. This detection rate is much higher than that of the
whole sample of cold IRAS sources in Sunada et al. (2007) where
222 out of 1563 sources was detected with maser emission, and further higher than those in
another two large surveys by Wouterloot et al. (1993) and Palumbo et al.
(1994). The detection rate of water masers can be explained in a framework of
the evolutionary sequence of MSF (Schnee \& Carpenter 2009): at an early
stage, the H\footnotesize\,II \normalsize regions are gravitationally bound to
the protostars and remain compact in size and can not be detected at centimeter
wavelength with current sensitivities; as protostars grow and become massive
enough to produce detectable H\footnotesize\,II \normalsize regions that can
not be suppressed by the stellar gravitational potential (Keto 2007); and then
the H\footnotesize\,II \normalsize regions expand and destroy the accretion
disks and vast parts of the ambient cloud, which are thought to be the source
of water masers and outflows. Water masers have not been found towards 14 out
of 39 IRAS sources and these sources may not be good HMPO candidates (see
Table~\ref{hmpo-table}).

\subsubsection{HMPO candidates identified in this sample}\label{hmpo-list}
An IRAS source is thought to contain a star cluster instead of a
single star. By adopting the universal initial mass function (IMF) of the stars
following Kroupa (2001) and the mass-luminosity relationship of the zero age
main sequence stars (Schaller et al. 1992), the mass of the most massive star
formed in the cluster is  8~\msun\, if the FIR luminosity of an IRAS source is
$\sim$8$\times$10$^3$~\lsun. Table~\ref{hmpo-table} presents the final HMPO
sample carefully identified after checking association between dense clumps and
IRAS sources in $\S$~\ref{dense}, the radio properties in $\S$~\ref{radio}, and
the existence of water masers in $\S$~\ref{water}. In total, we have identified
12 HMPO candidates in 27 sources with luminosities above
8$\times$10$^3$~\lsun.

\subsubsection{Subgroups in the sample}\label{subgroup}
To better investigate the physical conditions of MSFRs, we
divided the 39 sources into five subgroups based on their
luminosities and identified HMPO candidates as below: 1. subgroup {\bf Low}
with luminosities less than 8$\times~10^{3}$~\lsun\, and the mass of the most
massive star formed in the cluster is less than 8~\msun; 2. subgroup {\bf HMPO}
identified in $\S\ref{hmpo-list}$, with a FIR luminosity ranging from
8$\times~10^{3}$ to 1$\times~10^{5}$~\lsun; 3. subgroup {\bf Control} contains
the remaining sources in the same FIR luminosity range as subgroup {\bf HMPO}; 4.
subgroup {\bf Extreme} with FIR luminosities over 1$\times~10^{5}$~\lsun; and
5. Subgroup {\bf High} contains the {\bf HMPO}, {\bf Control} and {\bf Extreme}
sources with FIR luminosities over 8.0$\times10^3$~\lsun.

We compute the dynamical timescale, free-fall timescale, star formation
timescale and kinetic energy for the sample and list their median values for
each subgroup in Table~\ref{grouptable}. A dynamical timescale, $t_{\rm dyn}$,
is defined as $t_{\rm dyn}\,=\,2R/v$, where R is the radius of molecular clump
and $v$ the velocity. Observed line width and velocity dispersion can be
related to one dimensional velocity dispersion via $\sigma_{\rm v}\,=\,\Delta
V/2.355$ (Pan \& Padoan 2009) and $v^2\,=\,3\sigma_{\rm v}^2$. A free-fall
timescale, $t_{\rm ff}$, can be obtained by $t_{\rm
ff}\,=\,{\sqrt\frac{3\pi}{32G\rho}}$, where $G$ is the gravitational constant
and $\rho$ the gas density determined by the virial gas mass and radius of a
molecular clump. A star formation timescale is estimated by adopting a turbulent
core model for massive star formation (McKee \& Tan 2003). We calculate the
kinetic energy of a molecular clump by $E_{\rm kin}\,=\,\frac{1}{2}M_{\rm
VIR}v^2$. These three timescales are typically a few 10$^5$ yrs and the kinetic
energies range from 3.1$\times10^{45}$ to 6.9$\times10^{46}$~ergs.

\subsection{Comments on individual HMPO candidates}\label{comment}
Here we briefly describe the HMPO candidates with detailed information in the literatures.

\subsubsection{\rm IRAS 02232+6138}
In Fig.\,\ref{csmap}, we present the \cs\, integrated intensity map. The CS
centroid clearly differ from the IRAS source with a large offset of
$\sim$20$\arcsec$. Two outflows, with well separated bipolar lobes, are
obviously detected with an offset of $\sim$25$\arcsec$ from the IRAS source
(Zapata et al. 2011). They are close to the CS centroid where the most
blue-skewed spectrum is detected (Fig.\,\ref{profile}), showing the distinct
signature of infalling motion (Wu \& Evans 2003).  The hot molecular core (HMC)
of W3(OH) in IRAS 02232+6138 was well studied by Wyrowski et al. (1997, 1999),
and it coincides with the CS centroid and the location with the most
blue-skewed spectrum.

\subsubsection{\rm IRAS 02575+6017}
IRAS 02575+6017 is the eastern part of the W5-East H\footnotesize\,II
\normalsize region, which is the eastern part of the W5 H\footnotesize\,II
\normalsize region (Carpenter et al. 2000; Niwa et al. 2009). Bright-rimmed
clouds in the W5-East H\footnotesize\,II \normalsize region (Niwa et al. 2009)
are well studied to investigate the triggered star formation caused by UV
radiation from the central exciting star. The dense gas around IRAS 02575+6017
is compressed by the UV radiation or stellar wind from the exciting star in the
W5-East H\footnotesize\,II \normalsize region, which results in that the dense
cloud has a head-tail structure, as shown in Fig.\,\ref{csmap}. 

\subsubsection{\rm IRAS 06099+1800}\label{06099}
This region is a part of the S254-258 complex as a group of five
H\footnotesize\,II \normalsize regions. Most studies in the complex
focus on the region around IRAS\,06099+1800, S255IR and the region, S255N, 1$'$
north of S255IR (Miralles et al. 1997; Minier et al. 2005; Cyganowski et al.
2007; Klaassen \& Wilson 2007; Chavarr\'{\i}a et al. 2008). The former has a
young massive star cluster, and the latter is associated with signposts of
massive star formation at an early stage since no near infrared source is
associated with it.  Cyganowski et al. (2007) showed evidence of a massive
protocluster in S255N.  They found three cores at 1.3~mm with no IR
counterpart, with masses between 6 and 35\,\msun.

In Fig.\,\ref{csmap}, we present the \cs\, integrated intensity map around
IRAS\,06099+1800 and reveal that the IRAS source is deeply embedded in the
center of the dense gas cloud, which is elongated from north to south-east.
\cs\, emission (see Fig.\,\ref{csmap}) peaks at S255IR and elongated towards
S255N, but N$_2$H$^+$(1-0) peaks at S255N (Pirogov et al. 2003).  Klaassen \&
Wilson (2007) found the signatures of infall towards S255IR and S255N. Both of
the regions are HMPO candidates, but S255N is at the earlier stage.

\subsubsection{\rm IRAS 22543+6145}
In Fig.\,\ref{csmap}, we present the CS integrated intensity map and the
3.6\,cm VLA map around IRAS\,22543+6145. The dense gas is elongated from north
to south, and two dense molecular clumps are revealed, one is associated with
the IRAS source and another one is 140$\arcsec$ south of the IRAS source. Many
authors (e.g., Narayanan et al. 1996; Wu et al. 2005; Xu et al. 2006b) have
found the bipolar outflow towards this region. Wu \& Evans (2003) and Klaassen
\& Wilson (2007) found the signature of infall, which suggests that the IRAS
source could be a HMPO.

\subsubsection{\rm IRAS\,23033+5951}
As shown in Fig.\,\ref{csmap}, IRAS\,23033+5951 is deeply embedded in the
center of the dense molecular clump. Outflows in this region are
well studied by many authors (e.g., Beuther et al. 2002b; Reid \&
Matthews 2008). Reid \& Matthews (2008) also presented
high-resolution interferometric data and revealed two massive cores,
one with a massive protostar and another with the signature of
infall suggesting that it might be collapsing.

\subsubsection{\rm IRAS\,23385+6053}
In Fig.\,\ref{csmap}, IRAS\,23385+6053 is located in the center of the dense
molecular clump. The IRAS source is associated with a powerful bipolar outflow
seen in HCO$^{+}$ and SiO emission (Molinari et al. 1998b) and in CO J=2-1 (Wu
et al. 2005; Zhang et al. 2005). However, the most compact VLA configuration
revealed the existence of two nearby extended radio sources (Molinari et al.
2002), but no near infrared source is detected at the position of the IRAS
source (Thompson \& Macdonald 2003). Based on the physical and chemical nature
of IRAS\,23385+6053, Thompson \& Macdonald (2003) concluded that the IRAS
source is prior to the hot molecular core stage.

\section{Discussion}\label{discussion}

\subsection{Luminosity to mass ratio} \label{hmpo-ratio}

The ratio of infrared luminosity to molecular mass is roughly proportional to the star
formation rate per unit mass, and could be a good indicator for star formation
processes.  To investigate the evolution of MSFRs, we compare the infrared
luminosity to mass ratios of dense molecular clumps among different subgroups
in our sample. In general, there is a trend that more luminous sources have
higher \lmvir\, ratios as shown in Figure~\ref{lm}(left).  A natural
explanation for different \lmvir\, ratios is that the sources with lower ratios
are younger since they are at the earlier evolutionary stage and have not yet
reached their full luminosities (Sridharan et al. 2002).  Usually, most massive
stars are formed in most luminous clusters, and the massive stars dominate the
luminosity of the clusters according to the mass-to-luminosity relation for
zero age main sequence stars, $L\,\propto\,M^{\rm 3.5}$, and the IMF in the
cluster (Kroupa 2001). Therefore, the \lmvir\, ratios might be an ambiguous
indicator of the evolution of MSFRs: both massive stars formed in the cluster
and late evolutionary stage of MSFR can lead to high ratios. Fa\'{u}ndez et al.
(2004) argued that the \lmvir\, ratio could be interpreted as an indicator of
which type of most massive star formed in the clump rather than an evolutionary
effect.

To better understand the \lmvir\, ratio, we divided our sample
into five subgroups in $\S\ref{subgroup}$ and list their median
values in Table~\ref{grouptable}. The median \lmvir\, ratios are 18, 28, 110
and 310 \lsun/\msun\, for the subgroups {\bf Low}, {\bf HMPO}, {\bf Control}
and {\bf Extreme}, respectively. The ratios among {\bf Low}, {\bf Control} and
{\bf Extreme} sources increase with their luminosities, which can most likely
be interpreted as an indicator of which type of most massive star formed in the
cluster. However, the {\bf HMPO} sources have much lower \lmvir\, ratios than
the {\bf Control} ones, which is clearly seen Figure~\ref{lm}(left). This can
not be interpreted by the type of massive star formed in the cluster since both
types of sources share a similar luminosity range.  Indeed, the {\bf HMPO}
sources are still deeply embedded in the dense clumps with much higher
molecular masses compared to the {\bf Control} ones (see the middle panel of
Figure~\ref{lm}).  Half of the latter sources are not associated with dense
molecular clumps. We conclude that the {\bf HMPO} sources are in an early
evolutionary stage when the massive protostars have not reached their full
luminosities of the zero-age main sequence stars and are still accreting
material, supported by signatures of infall motions seen in about half of the
{\bf HMPO} sources (see the details in $\S\ref{comment}$). 
Giannetti et al. (2017) proposed that the \lm\, ratio is
directly related with the gas temperature of molecular gas in a simple toy
model of a spherical, internally heated clump, and the gas temperature is an
indicator of the evolutionary stage of the clump. They also pointed out that
most HMPOs have \lm\, ratios from 10~\lmsun\, to $\le$ 40~\lmsun, which is
consistent with the values of more than half identified HMPOs in our paper.

We also compare our sample to a subsample with potential
molecular outflows in the RMS survey (Lumsden et al. 2013; Maud et al.  2015a).
Figure~\ref{lm} shows the HMPOs (named MYSOs in Maud et al. 2015a) have
slightly higher \lmvir\, ratios than the compact \hii\, regions in the RMS
sample. It also happens to the HMPOs and \hii\, regions in the ATLASGAL survey
(Urquhart et al. 2014). As Maud et al. (2015a) pointed out, this may be due to
the fact that the HMPOs and compact \hii\, regions are belonging to a similar
evolutionary stage.  However, the typical \lmvir\, values of HMPOs in the RMS
sample are about two times as those in our sample. This may be due to a
systematic change on calculating the gas masses with different tracers at
different angular resolutions. The control sample in our paper includes much 
evolved sources, leading to this sample with high \lmvir\, ratios.

\subsection{Implication on massive star formation}

Having determined the physical properties of molecular clumps in MSFRs, we can
investigate the massive star formation process with the derived parameters. The
{\bf HMPO} sources have the largest sizes and the {\bf Low} have the smallest
ones. The line widths have large variance and the median values are 1.74,
3.00, 1.91 and 3.55 \kms\, for the subgroups {\bf Low}, {\bf HMPO}, {\bf
Control} and {\bf Extreme}, respectively.  Figure~\ref{lm}(right) presents
linewidth against FIR luminosity for our sample. The linewidth is an indicator
of turbulent motion of the clumps associated with the IRAS sources. In general
more luminous sources tend to be associated with clumps with broader line
widths, but the scatter is large.  Comparing the two subgroups {\bf HMPO} and
{\bf Control}, one can find that the former has a much broader linewidth than
the latter. The median linewidth of the {\bf Extreme} sources is the broadest
one among all subgroups (see Table~\ref{grouptable}). 

To investigate the impact of the feedback from MSFRs, we compare our results
with the outflow study by Zhang et al. (2005). Table~\ref{grouptable} lists
basic outflow parameters determined from Zhang's sample with a similar FIR
luminosity range as the subgroup {\bf High}.  We find that the typical kinetic
energy of the outflows is about half of the kinetic energy of dense clumps of
the {\bf High} sources and one sixth of the {\bf HMPO} sources, and the outflow
timescale is one fourth of the dynamical values of the {\bf High} and {\bf
HMPO} sources. It supports that the feedback (e.g., outflows) from MSFRs can
play an important role on the surrounding environments. The variance of kinetic
energies of different subgroups is quite large. The {\bf Low} sources have the
lowest kinetic energies since they only form low/intermediate mass stars.
Both {\bf HMPO} and {\bf Control} sources are forming massive stars.  However,
the {\bf Control} sources may have finished their mass accretion process, and
their central protostars have reached their zero-age main sequence stages with
full luminosities and strong UV radiation, ionizing and destroying the
surrounding molecular clouds. If no continuous energy supply, kinetic energy in
molecular clumps will dissipate as the sources evolve with time, and this may
result in narrower linewidths at the later stage as seen in the {\bf Control}
sources.  The newly formed massive stars in the {\bf Extreme} sources have
destroyed their natal clumps and the large linewidths might be due to the
strong stellar wind from the central massive stars. The {\bf HMPO} sources are
still in a relatively early evolutionary stage when they are deeply embedded in
the dense clumps with weak radio emission from the central protostars confined
in small regions. Their central protostars have high mass accretion rate and
are ejecting large amount of kinetic energy into the surrounding material
through jets and outflows. Their feedbacks are very important to maintain the
turbulent motion in the clumps as the median kinetic energy ratio of the
outflows to the clumps ranges from 40\% to 70\% if considering the different
timescales. Based on a much large outflow sample identified in
the ATLASGAL survey, Yang et al. (2018) also find that outflow energy is
comparable to the turbulent energy within the clump. It seems to support that
turbulence in the {\bf HMPO} sources may be largely supplied by the feedbacks
from MSFRs. However, they find no evidence that outflows contribute
significantly to the turbulent kinetic energy of surrounding materials. This is
also confirmed by another work in a follow-up study of the RMS survey (Maud et
al. 2015b). It suggests alternative means or undetected outflows may be
responsible for the turbulence in the clumps. 

\section{Conclusion}\label{conclusion}
We have mapped 39 massive star forming regions around the cold IRAS sources
with high-velocity CO flows in \cs.  All sources are detected in \cs, showing
the existence of CS clumps around the IRAS sources. We have identified 12
high-mass protostellar object (HMPO) candidates in this sample. The HMPO
candidates are characterized by low FIR luminosity to virial mass ratio because
they are at very early evolutionary stages of star formation when the massive
protostars have not reached their full luminosities of the zero-age main
sequence stars. Large turbulent motion in the HMPO sources may be
largely due to the large kinetic energy ejected by the central protostars
formed in dense clumps. However, alternative means or undetected outflows may also be
responsible for the turbulence in the clumps.

\begin{acknowledgements}

We thank the anonymous referee for valuable comments that improved this
	manuscript. We are grateful for supporting by the staff of Nobeyama
	Radio Observatory, Japan.  Y.A. acknowledges partial support by NSFC
	grant 11373007.

\end{acknowledgements}

\begin{figure*}[t]
\centering
  \includegraphics[angle=0,width=1.0\textwidth]{fig1.pdf}
  \caption{Mapping positions and profile map towards IRAS~02232+6138. The
spectra marked by the crosses with the same color are taken from one snapshot
observation, providing 25 spectra by the BEARS simultaneously. In total, four
snapshot observations and 100 spectra are obtained for this target.
The offsets are relative to the coordinate in Table~\ref{list}.}\label{profile}
\end{figure*}

\begin{figure*}[t]
\centering
  \includegraphics[angle=0,width=0.95\textwidth]{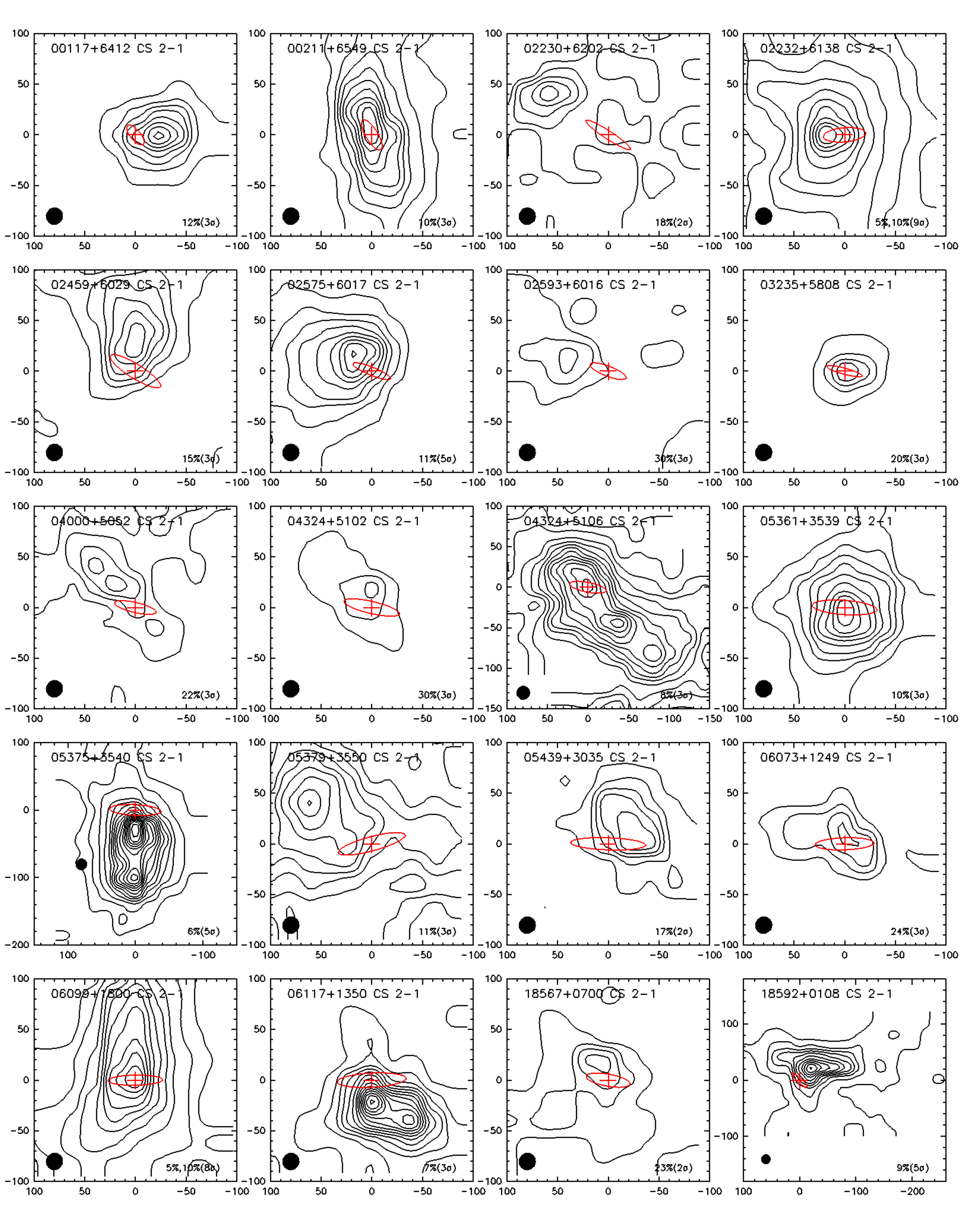}
  \caption{Contour maps of \cs\, integrated intensity.
  The source name and half power beam of NRO45m are give in the upper and
  lower-left of each panel. The contour level is given at the lower-right of each panel.
  For instance, "5\%,10\%(9$\sigma$)" means the first contour is 5\% of the peak intensity,
  the next contour is 10\% of the peak intensity, and the interval is 10\% or 9$\sigma$.
  The positions and the error ellipse of IRAS sources are also marked with 
  plus sign and ellipse. The positions of IRAS point sources are at (0,0).
}\label{csmap}
\end{figure*}

\addtocounter{figure}{-1}
\begin{figure*}
  \includegraphics[angle=0,width=0.95\textwidth]{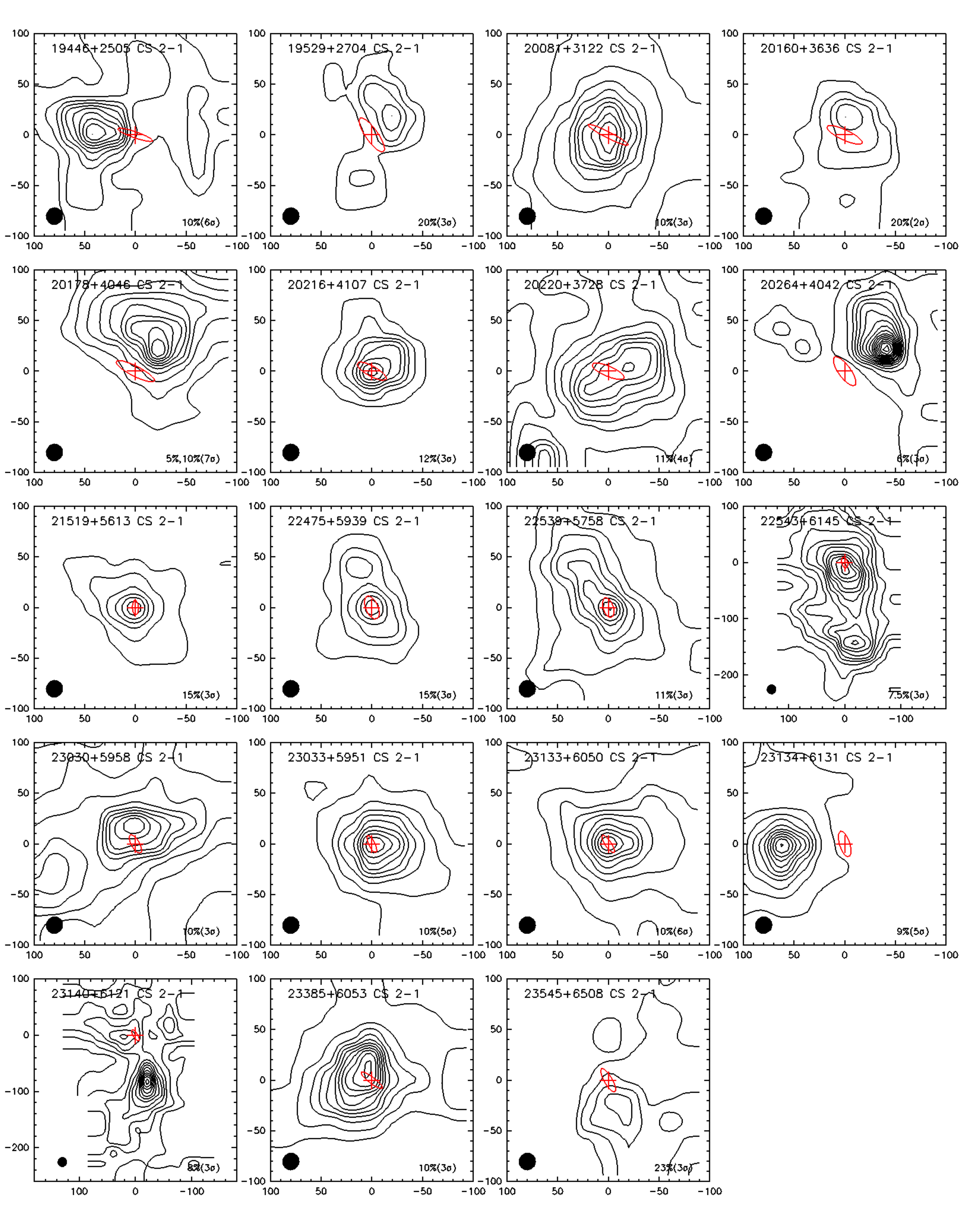}
  \caption{\small continued.}
\end{figure*}

\begin{figure*}[t]
\centering
\includegraphics[angle=0,width=1\textwidth]{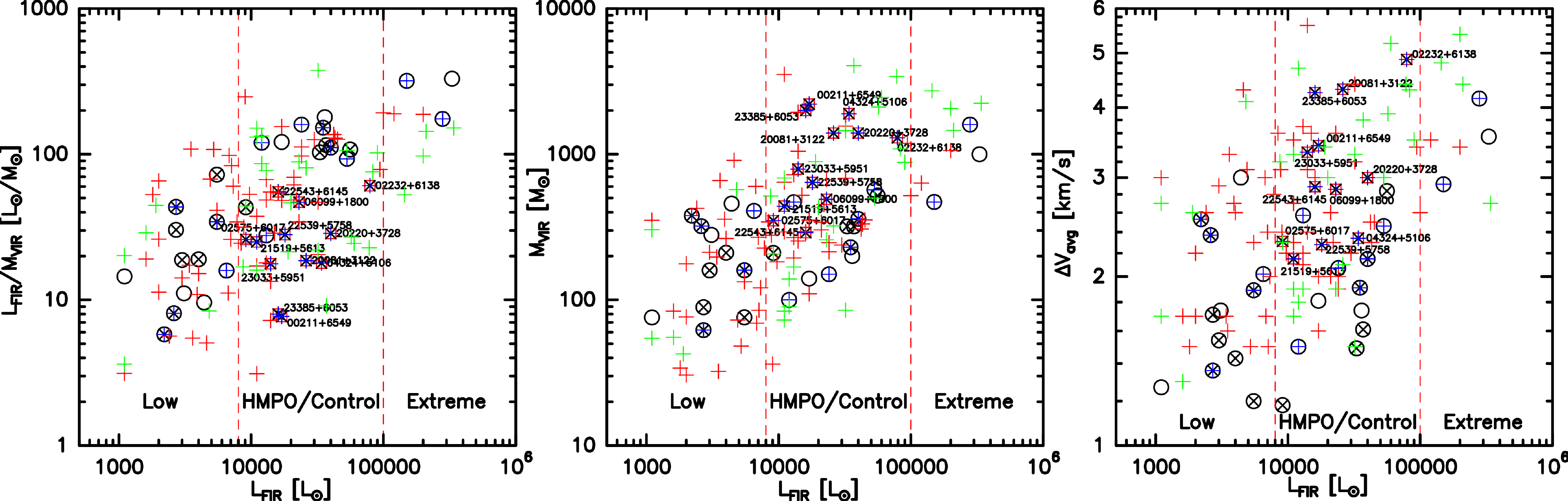}
\caption{Luminosity plotted against Luminosity-to-mass ratio {\bf (left
	panel)}, virial mass {\bf (middle panel)}, and line width {\bf (right
	panel)}.  The sources with cross symbols are deeply embedded in the
	dense clumps, and the ones with blue plus symbols associated with water
	masers. The red squares denote HMPO candidates and the open circles are
	the remaining sources. The YSOs and \hii\, regions in Maud et al.
	(2015a), where the virial masses are derived from C$^{\rm 18}$O
	($J$\,=\,3$\to$2) data, are also presented as red and green plus signs,
	respectively.  Vertical dashed lines in red represents the luminosities
	of $\sim8\times10^3$ \lsun\, (above this threshold, the dense clumps
	will form massive stars) and $1.0\times10^5$ \lsun.}\label{lm}
\end{figure*}

\begin{center}
\begin{table*}[t]
\centering
\caption{The source list}\label{list}
\begin{tabular}{cccccccccc}
\hline
     IRAS Name    &      RA
&    Dec. &    $L_{\rm IR}$ &  $\rm Dist.$   & Dist.   &  Comment$^a$ \\
                &      (J2000.0) 
&    (J2000.0)  &     (\lsun)  &  (kpc)  &   ref. \\
\hline
     00117+6412 & 00:14:27.7 & +64:28:46 &    2.2E+03 &  2.57\ppm0.25 & 1   & kinematic distance    \\ 
     00211+6549 & 00:23:58.0 & +66:06:03 &    1.7E+04 &  5.60\ppm0.33 & 1   & kinematic distance   \\ 
     02230+6202 & 02:26:50.8 & +62:15:52 &    2.4E+04 &  1.95\ppm0.04 & 2   & parallax measurement   \\ 
     02232+6138 & 02:27:01.0 & +61:52:13 &    7.9E+04 &  1.95\ppm0.04 & 2   & parallax measurement    \\ 
     02459+6029 & 02:49:47.6 & +60:42:07 &    3.0E+03 &  1.95\ppm0.04 & 2   & parallax measurement    \\ 
     02575+6017 & 03:01:29.2 & +60:29:12 &    9.1E+03 &  1.95\ppm0.04 & 2   & parallax measurement    \\ 
     02593+6016 & 03:03:17.8 & +60:27:52 &    1.2E+04 &  1.95\ppm0.04 & 2   & parallax measurement   \\ 
     03235+5808 & 03:27:31.1 & +58:19:21 &    2.7E+03 &  1.95\ppm0.04 & 2   & parallax measurement    \\ 
     04000+5052 & 04:03:49.3 & +51:00:48 &    1.1E+03 &  1.5$^b$ & 3   & photometric distance    \\ 
     04324+5102 & 04:36:16.0 & +51:08:12 &    9.1E+03 &  6.0\ppm0.6 & 4   & photometric distance    \\ 
     04324+5106 & 04:36:19.6 & +51:12:44 &    3.4E+04 &  6.0\ppm0.6 & 4   & photometric distance    \\ 
     05361+3539 & 05:39:27.6 & +35:40:42 &    2.6E+03 &  1.8$^b$ & 5   & photometric distance    \\ 
     05375+3540 & 05:40:53.6 & +35:42:15 &    1.3E+04 &  1.8$^b$ & 5   & photometric distance    \\ 
     05379+3550 & 05:41:20.5 & +35:52:06 &    3.1E+03 &  1.8$^b$ & 5   & photometric distance    \\ 
     05439+3035 & 05:47:12.8 & +30:36:11 &    4.0E+03 &  3.5\ppm0.5 & 6   & photometric distance    \\ 
	06073+1249 & 06:10:12.4 & +12:48:45 &    3.3E+04 &  4.69$^{+0.95}_{-0.83}$ & 1   & kinematic distance    \\ 
     06099+1800 & 06:12:53.3 & +17:59:22 &    2.3E+04 &  1.59$^{+0.07}_{-0.06}$ & 7   & parallax measurement    \\ 
     06117+1350 & 06:14:36.5 & +13:49:35 &    5.3E+04 &  3.8\ppm1.0 & 4   & photometric distance    \\ 
     18567+0700 & 18:59:13.5 & +07:04:47 &    5.5E+03 &  2.09\ppm0.18 & 1   & near kinematic distance    \\ 
     18592+0108 & 19:01:46.9 & +01:13:07 &    2.8E+05 &  3.27$^{+0.56}_{-0.42}$ & 8   & parallax measurement    \\ 
     19446+2505 & 19:46:47.0 & +25:12:43 &    1.5E+05 &  2.5\ppm0.4 & 9   & photometric distance    \\ 
     19529+2704 & 19:54:59.6 & +27:12:52 &    3.6E+04 &  3.2\ppm1.1 & 10   & photometric distance    \\ 
     20081+3122 & 20:10:09.1 & +31:31:34 &    2.6E+04 &  2.57$^{+0.34}_{-0.27}$ & 7   & parallax measurement    \\ 
     20160+3636 & 20:17:56.1 & +36:45:33 &    3.7E+04 &  4.4\ppm1.4 & 4   & photometric distance    \\ 
     20178+4046 & 20:19:39.2 & +40:56:30 &    1.7E+04 &  1.68\ppm1.32 &  1  & kinematic distance    \\ 
     20216+4107 & 20:23:23.8 & +41:17:39 &    5.5E+03 &  2.35\ppm0.77 &  1  & far kinematic distance    \\ 
     20220+3728 & 20:23:55.6 & +37:38:10 &    4.0E+04 &  3.68\ppm0.47 &  1  & far kinematic distance    \\ 
     20264+4042 & 20:28:12.4 & +40:52:27 &    3.3E+05 &  7.09\ppm0.27 &  1  & kinematic distance    \\ 
     21519+5613 & 21:53:39.2 & +56:27:45 &    1.1E+04 &  6.01\ppm0.28 &  1  & kinematic distance   \\ 
     22475+5939 & 22:49:29.4 & +59:54:56 &    4.0E+04 &  4.11\ppm0.27 &  1  & kinematic distance    \\ 
     22539+5758 & 22:56:00.0 & +58:14:45 &    1.8E+04 &  4.56\ppm0.27 &  1  & kinematic distance    \\ 
     22543+6145 & 22:56:19.1 & +62:01:57 &    1.6E+04 &  0.70\ppm0.04 &  11  & parallax measurement    \\ 
     23030+5958 & 23:05:10.6 & +60:14:40 &    3.5E+04 &  2.87\ppm0.75 &  12  & photometric distance    \\ 
     23033+5951 & 23:05:25.1 & +60:08:11 &    1.4E+04 &  2.87\ppm0.75 &  12  & photometric distance   \\ 
     23133+6050 & 23:15:31.4 & +61:07:09 &    5.6E+04 &  2.97\ppm0.31 &  12  & photometric distance    \\ 
     23134+6131 & 23:15:34.6 & +61:47:42 &    4.4E+03 &  2.8\ppm0.9 &  4  & photometric distance    \\ 
     23140+6121 & 23:16:11.7 & +61:37:44 &    6.5E+03 &  2.8\ppm0.9 &  4  & photometric distance    \\ 
     23385+6053 & 23:40:53.1 & +61:10:21 &    1.6E+04 &  3.86\ppm0.27 &  1  &  kinematic distance   \\ 
     23545+6508 & 23:57:05.2 & +65:25:10 &    2.7E+03 &  1.14\ppm0.24 &  1  &  kinematic distance   \\ 

\hline
\end{tabular}
\begin{list}{}{}

\item{$^{\mathrm{a}}$ Kinematic distances are calculated using the revised
prescription given in Reid et al. (2009).  A near kinematic distance is adopted
for 18567+0700 and far kinematic distances for 20216+4107 and 20220+3728.}
\item{$^{\mathrm{b}}$ For these sources, errors are not assigned in the literature.}
\item{$References:$ (1) This work (2) Xu et al. (2006a);
	(3) Esteban et al. (2004); (4) Blitz \& Fich (1982); (5) Evans \& Blair (1981);
(6) Tapia et al. (1997); (7) Rygl et al. (2010); (8) Zhang et al. (2009);
(9) Marin-Franch et al. (2009); (10) Fich et al. (1989); 
(11) Moscadelli et al. (2009); (12) Russeil et al. (2007).
}
\end{list}
\end{table*}
\end{center}

\begin{center}
\begin{table*}[t]
\centering
\caption{Observational parameters of peak \cs\, spectra
for IRAS sources}\label{spectable}
\small
\begin{tabular}{ccccccc}
\hline
 IRAS Name  &  $\Delta$$\alpha$  &  $\Delta$$\delta$  & $T_{\rm mb}$  &  $\rm
V_{LSR}$  &  Line Width  &  $\int T_{\rm mb}{\rm dv}$  \\
   &  ($\arcsec$)  &  ($\arcsec$)  & (K)  &  ($\kms$)  &  ($\kms$)  &  (K~$\kms$) \\
\hline
  00117+6412& -20.6&   0.0&    3.05(40)&   -36.2&    2.44(12)&    7.44(30) \\
  00211+6549&   0.0&   0.0&    3.93(49)&   -68.8&    3.03(13)&   11.90(41) \\
  02230+6202&  61.5&  41.1&    2.11(58)&   -42.2&    2.15(22)&    4.55(40) \\
  02232+6138&  20.5&   0.0&   11.22(64)&   -47.2&    6.41( 8)&   71.88(76) \\
  02459+6029&   0.0&  41.1&    3.30(38)&   -36.8&    1.52( 8)&    5.01(22) \\
&   0.0&  41.1&    2.04(38)&   -40.3&    1.29(15)&    2.64(22) \\
  02575+6017&  20.5&  20.5&    7.32(53)&   -37.7&    2.47( 6)&   18.07(39) \\
  02593+6016&  41.1&  20.5&    2.44(66)&   -37.7&    1.73(19)&    4.23(40) \\
  03235+5808&   0.0&   0.0&    3.36(62)&   -46.5&    1.84(13)&    6.18(39) \\
  04000+5052&  41.1&  41.1&    2.25(40)&   -31.4&    1.21(11)&    2.72(21) \\
  04324+5102&   0.0&   0.0&    1.64(51)&   -36.5&    2.36(30)&    3.88(39) \\
  04324+5106&   0.0&   0.0&    3.12(27)&   -36.4&    2.89( 8)&    9.01(22) \\
  05361+3539&   0.0&   0.0&    5.57(65)&   -16.7&    2.54(13)&   14.16(54) \\
  05375+3540&   0.0& -20.6&   10.52(44)&   -17.0&    2.64( 4)&   27.76(34) \\
  05379+3550&  61.6&  41.1&    4.30(43)&   -21.0&    1.73( 8)&    7.44(27) \\
  05439+3035& -20.6&   0.0&    1.86(40)&   -17.9&    1.51(14)&    2.82(23) \\
  06073+1249&   0.0&   0.0&    2.39(56)&    25.5&    1.80(18)&    4.30(36) \\
  06099+1800&   0.0&   0.0&   16.65(81)&     7.5&    3.03( 5)&   50.37(67) \\
  06117+1350&   0.0& -20.6&    5.55(44)&    17.7&    2.72( 7)&   15.10(34) \\
  18567+0700&  20.5&  20.5&    3.00(94)&    29.6&    1.68(26)&    5.06(61) \\
  18592+0108& -20.6&  20.5&    7.35(68)&    43.4&    4.97(11)&   36.48(70) \\
  19446+2505&  41.1&   0.0&   11.36(62)&    21.3&    3.06( 6)&   34.72(52) \\
  19529+2704& -20.6&  20.5&    4.86(85)&    21.9&    1.76(13)&    8.54(53) \\
  20081+3122&   0.0&   0.0&    5.00(72)&    11.6&    4.81(16)&   24.05(72) \\
  20160+3636&   0.0&  20.5&    2.65(72)&    -0.0&    1.62(25)&    4.31(48) \\
  20178+4046& -20.6&  20.5&   10.05(41)&     1.0&    1.89( 3)&   19.03(26) \\
  20216+4107&   0.0&   0.0&    4.81(61)&    -1.5&    2.14(11)&   10.29(43) \\
  20220+3728& -20.6&   0.0&    4.14(44)&    -2.6&    3.55(12)&   14.68(40) \\
  20264+4042& -41.1&  20.5&    6.50(49)&   -41.1&    3.62( 8)&   23.57(43) \\
  21519+5613&   0.0&   0.0&    2.65(46)&   -63.0&    2.63(18)&    6.97(37) \\
  22475+5939&   0.0&   0.0&    2.28(55)&   -49.1&    5.81(27)&   13.24(59) \\
  22539+5758&   0.0&   0.0&    4.61(56)&   -54.2&    2.37(10)&   10.93(40) \\
  22543+6145&   0.0& -20.6&    5.89(52)&   -11.5&    4.32(11)&   25.44(52) \\
  22543+6145S& -20.6&-143.9&    2.89(38)&   -10.0&    6.54(16)&   18.89(44) \\
  23030+5958&  20.5&   0.0&    2.92(64)&   -52.9&    1.84(17)&    5.36(43) \\
&  20.5&   0.0&    7.47(64)&   -49.9&    1.76( 7)&   13.13(44) \\
  23033+5951&   0.0&   0.0&    6.94(60)&   -52.9&    3.51(10)&   24.35(54) \\
  23133+6050&   0.0&   0.0&   11.84(69)&   -56.3&    2.67( 5)&   31.67(54) \\
  23134+6131&  61.6&   0.0&    8.00(61)&   -48.7&    3.35( 9)&   26.79(54) \\
  23140+6121&  20.5&   0.0&    2.69(47)&   -51.5&    2.26(15)&    6.08(34) \\
  23140+6121S& -20.6& -82.2&    3.51(25)&   -49.8&    3.78( 8)&   13.23(23) \\
  23385+6053&   0.0&   0.0&    3.21(48)&   -49.6&    3.76(16)&   12.04(44) \\
  23545+6508& -20.6& -20.6&    2.22(42)&   -18.2&    1.35(14)&    2.99(24) \\
\hline
\end{tabular}
\end{table*}
\end{center}

\begin{center}
\begin{table*}[h]
\centering
\caption{Derived parameters of \cs\, maps for IRAS
sources}\label{core}
\begin{tabular}{cccccc}
\hline
  IRAS Name &  $R$   &  $\rm \overline{\Delta v} $  & $M_{\rm VIR}$  &  \lmvir & Notes$^{\rm a}$   \\
 &  (pc)   &  ($\kms$)  & (\msun)   &  (\lmsun)  &  \\

\hline
     00117+6412 &  0.29 &  2.53 &    3.8E+02 &    5.8E+00 &    Y \\ 
     00211+6549 &  0.91 &  3.42 &    2.2E+03 &    7.6E+00 &    Y \\ 
     02230+6202 &  0.17 &  2.07 &    1.5E+02 &    1.6E+02 &    N \\ 
     02232+6138 &  0.26 &  4.87 &    1.3E+03 &    6.0E+01 &    Y \\ 
     02459+6029 &  0.33 &  1.54 &    1.6E+02 &    1.8E+01 &    Y \\ 
     02575+6017 &  0.31 &  2.31 &    3.5E+02 &    2.6E+01 &    Y \\ 
     02593+6016 &  0.21 &  1.50 &    1.0E+02 &    1.1E+02 &    N \\ 
     03235+5808 &  0.15 &  1.71 &    8.9E+01 &    3.0E+01 &    Y \\ 
     04000+5052 &  0.23 &  1.27 &    7.6E+01 &    1.5E+01 &    N \\ 
     04324+5102 &  0.73 &  1.18 &    2.1E+02 &    7.8E+01 &    Y \\ 
     04324+5106 &  1.66 &  2.34 &    1.9E+03 &    3.2E+01 &    Y \\ 
     05361+3539 &  0.27 &  2.37 &    3.2E+02 &    8.0E+00 &    Y \\ 
     05375+3540 &  0.34 &  2.57 &    4.7E+02 &    2.9E+01 &    N \\ 
     05379+3550 &  0.44 &  1.74 &    2.8E+02 &    1.1E+01 &    N \\ 
     05439+3035 &  0.49 &  1.43 &    2.1E+02 &    1.9E+01 &    Y \\ 
     06073+1249 &  0.68 &  1.49 &    3.2E+02 &    1.0E+02 &    Y \\ 
     06099+1800 &  0.29 &  2.86 &    4.9E+02 &    4.7E+01 &    Y \\ 
     06117+1350 &  0.45 &  2.46 &    5.7E+02 &    9.3E+01 &    N \\ 
     18567+0700 &  0.25 &  1.20 &    7.6E+01 &    7.2E+01 &    Y \\ 
     18592+0108 &  0.45 &  4.15 &    1.6E+03 &    1.7E+02 &    N \\ 
     19446+2505 &  0.26 &  2.92 &    4.7E+02 &    3.1E+02 &    N \\ 
     19529+2704 &  0.32 &  1.74 &    2.0E+02 &    1.8E+02 &    N \\ 
     20081+3122 &  0.37 &  4.31 &    1.4E+03 &    1.8E+01 &    Y \\ 
     20160+3636 &  0.59 &  1.61 &    3.2E+02 &    1.2E+02 &    Y \\ 
     20178+4046 &  0.21 &  1.81 &    1.4E+02 &    1.1E+02 &    N \\ 
     20216+4107 &  0.22 &  1.89 &    1.6E+02 &    3.4E+01 &    Y \\ 
     20220+3728 &  0.76 &  3.00 &    1.4E+03 &    2.8E+01 &    Y \\ 
     20264+4042 &  0.39 &  3.55 &    1.0E+03 &    3.2E+02 &    N \\ 
     21519+5613 &  0.46 &  2.15 &    4.4E+02 &    2.5E+01 &    Y \\ 
     22475+5939 &  0.37 &  2.15 &    3.6E+02 &    1.1E+02 &    Y \\ 
     22539+5758 &  0.59 &  2.28 &    6.4E+02 &    2.9E+01 &    Y \\ 
     22543+6145 &  0.17 &  2.89 &    2.9E+02 &    5.3E+01 &    Y \\ 
     23030+5958 &  0.30 &  1.91 &    2.3E+02 &    1.5E+02 &    Y \\ 
     23033+5951 &  0.33 &  3.33 &    7.9E+02 &    1.8E+01 &    Y \\ 
     23133+6050 &  0.33 &  2.84 &    5.2E+02 &    9.6E+01 &    Y \\ 
     23134+6131 &  0.24 &  3.00 &    4.6E+02 &    9.5E+00 &    N \\ 
     23140+6121 &  0.48 &  2.02 &    4.1E+02 &    1.6E+01 &    N \\ 
     23385+6053 &  0.52 &  4.25 &    2.0E+03 &    8.4E+00 &    Y \\ 
     23545+6508 &  0.16 &  1.36 &    6.2E+01 &    4.3E+01 &    Y \\ 
\hline
\end{tabular}
\begin{list}{}{}
\item{$^{\mathrm{a}}$ Y: an IRAS source, located within the contour
level of 50\% of the peak, is deeply embedded in the dense gas cloud;
N: an IRAS source is not deeply embedded in the dense gas cloud.}
\end{list}
\end{table*}
\end{center}

\begin{center}
\begin{table*}[t]
\centering
\caption{HMPO candidates in the sample}\label{hmpo-table}
\footnotesize
\begin{tabular}{cccccccccccc}
\hline
           & \multicolumn{3}{c}{FIR Properties}  & & \multicolumn{3}{c}{Radio Properties}  & \\
\cline{2-4}   \cline{6-8}  \\ 
     IRAS    &  $L_{\rm IR}$  & Mu$^a$  & Spec.$^b$  & & S$_{\rm
1.4GHz}$$^c$ &   ${\rm log  N{\rm _{c}^{'}}}$  & Spec.$^d$  &
H$_2$O$^e$  & Dense$^f$  & Massive$^g$  &
HMPO  \\
     Name    &  (\lsun)  &  (\msun)  &  Type  & &  (mJy)  &  ($\rm s^{-1}$)  &
Type      &   maser  &  Core   &  Star   &  Candidate \\

\hline
  00117+6412 &    2.2E+03 &    5.5 &    $>$B3 & &     $\le$2.5E+00 &   44.8 &   $\ge$B2 &    Y &    Y &    N &      N \\
  {\bf 00211+6549} &    1.7E+04 &   10.1 &     B1 &  &    $\le$2.5E+00 &   45.5 &   $\ge$B1 &    Y &    Y &    Y &      Y \\
  02230+6202 &    2.4E+04 &   11.3 &     B1 &  &    1.1E+04 &   48.2 &     O8 &    Y &    N &    N &      N \\
  {\bf 02232+6138} &    7.9E+04 &   16.7 &     B0 &  &    3.8E+01 &   45.7 &     B1 &    Y &    Y &    Y &      Y \\
  02459+6029 &    3.0E+03 &    6.0 &    $>$B3 & &      $\le$2.5E+00 &   44.6 &   $\ge$B2 &    N &    Y &    N &      N \\
  {\bf 02575+6017} &    9.1E+03 &    8.3 &     B2 &  &    $\le$2.5E+00 &   44.6 &   $\ge$B2 &    Y &    Y &    Y &      Y \\
  02593+6016 &    1.2E+04 &    8.9 &     B2 &  &    1.0E+03 &   47.2 &     B0 &    Y &    N &    N &      N \\
  03235+5808 &    2.7E+03 &    5.8 &    $>$B3 & &     2.0E+01 &   45.5 &     B1 &    N &    Y &    N &      N \\
  04000+5052 &    1.1E+03 &    4.6 &    $>$B3 & &     1.5E+01 &   45.1 &     B1 &    N &    N &    N &      N \\
  04324+5102 &    9.1E+03 &    8.3 &     B2 & &     $\le$2.5E+00 &   45.5 &   $\ge$B1 &    N &    Y &    Y &      N \\
  {\bf 04324+5106} &    3.4E+04 &   12.6 &   B0.5 &  &    2.0E+02 &   47.5 &     B0 &    Y &    Y &    Y &     Y \\
  05361+3539 &    2.6E+03 &    5.8 &    $>$B3 & &     $\le$2.5E+00 &   44.5 &   $\ge$B2 &    Y &    Y &    N &      N \\
  05375+3540 &    1.3E+04 &    9.3 &     B2 &  &    2.5E+02 &   46.5 &   B0.5 &    Y &    N &    N &      N \\
  05379+3550 &    3.1E+03 &    6.0 &     B3 &  &    $\le$2.5E+00 &   44.5 &   $\ge$B2 &    N &    N &    N &      N \\
  05439+3035 &    4.0E+03 &    6.5 &     B3 &  &    $\le$2.5E+00 &   45.1 &   $\ge$B1 &    N &    Y &    N &      N \\
  06073+1249 &    3.3E+04 &   12.4 &   B0.5 &  &    3.5E+01 &   46.5 &   B0.5 &    N &    Y &    Y &      N \\
  {\bf 06099+1800} &    2.3E+04 &   11.1 &     B1 &  &    $\le$2.5E+00 &   44.4 &   $\ge$B2 &    Y &    Y &    Y &      Y \\
  06117+1350 &    5.3E+04 &   14.7 &   B0.5 & &     2.3E+01 &   46.1 &   B0.5 &    Y &    N &    N &      N \\
  18567+0700 &    5.5E+03 &    7.2 &     B3 & &     $\le$2.5E+00 &   44.6 &   $\ge$B2 &    N &    Y &    N &      N \\
  18592+0108 &    2.8E+05 &   19.0 &   O9.5 & &     1.0E+04 &   48.6 &     O8 &    Y &    N &    N &      N \\
  19446+2505 &    1.5E+05 &   18.2 &     B0 &  &    4.0E+03 &   48.0 &   O8.5 &    Y &    N &    N &      N \\
  19529+2704 &    3.6E+04 &   12.9 &   B0.5 &  &    1.8E+03 &   47.9 &     O9 &    N &    N &    N &      N \\
  {\bf 20081+3122} &    2.6E+04 &   11.6 &     B1 &  &    3.4E+01 &   45.9 &   B0.5 &    Y &    Y &    Y &     Y \\
  20160+3636 &    3.7E+04 &   13.0 &   B0.5 &  &    4.5E+01 &   46.5 &   B0.5 &    N &    Y &    Y &      N \\
  20178+4046 &    1.7E+04 &   10.1 &     B1 &  &    2.8E+01 &   45.5 &     B1 &    N &    N &    N &      N \\
  20216+4107 &    5.5E+03 &    7.1 &     B3 &  &    $\le$2.5E+00 &   44.7 &   $\ge$B2 &    Y &    Y &    N &      N \\
  {\bf 20220+3728} &    4.0E+04 &   13.3 &   B0.5 &  &    6.3E+02 &   47.5 &     B0 &    Y &    Y &    Y &     Y \\
  20264+4042 &    3.3E+05 &   19.1 &   O9.5 &  &    4.0E+03 &   48.9 &     O8 &    N &    N &    N &      N \\
  {\bf 21519+5613} &    1.1E+04 &    8.8 &     B2 &  &    $\le$2.5E+00 &   45.5 &   $\ge$B1 &    Y &    Y &    Y &     Y \\
  22475+5939 &    4.0E+04 &   13.4 &   B0.5 &  &    2.3E+03 &   48.2 &     O8 &    Y &    Y &    Y &    N \\
  {\bf 22539+5758} &    1.8E+04 &   10.3 &     B1 &  &    $\le$2.5E+00 &   45.3 &   $\ge$B1 &    Y &    Y &    Y &      Y \\
  {\bf 22543+6145} &    1.6E+04 &    9.8 &     B1 &  &    5.9E+01 &   45.1 &     B1 &    Y &    Y &    Y &      Y \\
  23030+5958 &    3.5E+04 &   12.8 &   B0.5 &  &    1.9E+03 &   47.8 &     O9 &    Y &    Y &    Y &    N \\
  {\bf 23033+5951} &    1.4E+04 &    9.6 &     B2 &  &    $\le$2.5E+00 &   44.9 &   $\ge$B2 &    Y &    Y &    Y &      Y \\
  23133+6050 &    5.6E+04 &   14.7 &   B0.5 &  &    7.1E+02 &   47.4 &     B0 &    N &    Y &    Y &      N \\
  23134+6131 &    4.4E+03 &    6.7 &     B3 &  &    $\le$2.5E+00 &   44.9 &   $\ge$B2 &    N &    N &    N &      N \\
  23140+6121 &    6.5E+03 &    7.5 &     B3 &  &    $\le$2.5E+00 &   44.9 &   $\ge$B2 &    Y &    N &    N &      N \\
  {\bf 23385+6053} &    1.6E+04 &    9.9 &     B1 &  &    7.2E+01 &   46.6 &   B0.5 &    Y &    Y &    Y &     Y \\
  23545+6508 &    2.7E+03 &    5.8 &    $>$B3 & &     $\le$2.5E+00 &   44.1 &   $\ge$B3 &    Y &    Y &    N &      N \\
\hline
\end{tabular}
\begin{list}{}{}
\item{The HMPO candidates are shown in bold font.}
\item{$^{\mathrm{a}}$ The mass of the most massive star formed in a cluster.}
\item{$^{\mathrm{b}}$ Spectral type is determined by the luminosity of the most massive
star formed in a cluster (Panagia 1973).} 
\item{$^{\mathrm{c}}$ A upper limit of 5$\sigma$ is adopted for an undetected source.} 
\item{$^{\mathrm{d}}$ Spectral type is determined by the number of photons of the Lyman
continuum (Panagia 1973).} 
\item{$^{\mathrm{e}}$ Association with a detected water maser (Sunada et al. 2007).}
\item{$^{\mathrm{f}}$ Same as in Table~\ref{core}.}
\item{$^{\mathrm{g}}$ Whether the mass of the most massive star in a cluster is higher than 8~\msun.}
\end{list}
\end{table*}
\end{center}


\begin{center}
\begin{table*}[t]
\centering
\caption{Statistics of derived parameters for the subgroups of the sample}\label{grouptable}
\footnotesize
\begin{tabular}{ccccccccccc}
\hline
subgroup$^a$ &  Count$^b$ &  $L_{\rm FIR}$  &  $M_{\rm VIR}$  
& \lmvir & $\Delta V$ &  R  &  $t_{\rm dyn}$$^c$ & $t_{\rm ff}$$^c$ & $t_{\rm msf}$$^c$ & $E_{\rm kin}$$^c$  \\
 &  &  (\lsun)  &  (\msun)  & (\lmsun) & (\kms) &  (pc)  &  (yrs)   & (yrs)   &   (yrs)  &   (ergs)  \\
\hline

Low & 12/ 8/ 5 &  3.1E+03 &  2.1E+02 &  1.8E+01 &   1.74 &   0.27 &    4.8E+05 &    2.1E+05 &    ... &    3.1E+45 \\
HMPO & 12/12/12 &  1.8E+04 &  1.0E+03 &  2.8E+01 &   3.00 &   0.46 &    3.6E+05 &    1.5E+05 &    2.0E+05 &    4.6E+46 \\
Control & 12/ 6/ 6 &  3.5E+04 &  3.2E+02 &  1.1E+02 &   1.91 &   0.34 &    4.6E+05 &    2.0E+05 &    ... &    4.4E+45 \\
Extreme &  3/ 0/ 2 &  2.8E+05 &  1.0E+03 &  3.1E+02 &   3.55 &   0.39 &    2.9E+05 &    1.2E+05 &    ... &    6.9E+46 \\
High & 27/18/20 &  2.6E+04 &  4.8E+02 &  7.8E+01 &   2.46 &   0.37 &    3.6E+05 &    1.5E+05 &  ... &    1.8E+46 \\
Maud15a HMPOs$^d$ & 31 & 1.7E+04 & 3.4E+02 & 5.3E+01  & 2.5 & 0.44  &  & & & \\
Maud15a H\scriptsize\,II\footnotesize$^d$  & 21 & 2.3E+04 & 5.0E+02 & 7.6E+01  & 3.2 & 0.47 & & & & \\
Zhang05$^e$ & 21  &  2.4E+04 &  ... &  ... &   ... &   ... &    8.3E+04 &    ... &    ... &   7.2E+45   \\ 
\hline
\end{tabular}
\begin{list}{}{}
\item{$^{\mathrm{a}}$ The sample is divided into five subgroups (subgroups {\bf Low},
{\bf HMPO}, {\bf Control}, {\bf Extreme} and {\bf High}, see details in $\S~\ref{subgroup}$).}
\item{$^{\mathrm{b}}$ n1/n2/n3: n1 is the number of the sources in the subgroup,
n2 the number of the sources associated with dense clumps and n3 the number of
the sources with the detection of water masers.}
\item{$^{\mathrm{c}}$ See $\S~\ref{subgroup}$ for the details of the calculations
of timescales and energy. The star formation timescales are estimated only for
the {\bf HMPO} sources as their molecular clouds where the massive protostars are
forming are not largely affected by the strong feedbacks of MSFRs.}
\item{$^{\mathrm{d}}$ The selected HMPOs and compact H\scriptsize\,II\footnotesize\, regions in the RMS
	survey (Maud et al. 2015a) have a luminosity range from 8$\times$10$^3$ to 10$^5$ \lsun.}
\item{$^{\mathrm{e}}$ The median values (dynamical timescale and kinetic energy)
of outflows for the subsample in Zhang et al. (2005). This subsample share the
similar luminosity range as subgroup {\bf High}.}
\end{list}
\end{table*}
\end{center}

\end{document}